# Robust Broadband Infrared Unidirectional Absorption Enabled by a Non-Hermitian Multilayer

*Ayaha Yamamoto[A], Ganbat Batorgil[A], Yu-Jung Lu[B], Satoshi Iwamoto[C], and Wakana Kubo[A]**

A. Ayaha Yamamoto, Ganbat Batorgil, and Wakana Kubo
Department of Electrical Engineering and Computer Science, Tokyo University of Agriculture and Technology, 2-24-16 Naka-cho, Koganei-shi, Tokyo 184-8588, Japan.
E-mail: w-kubo@cc.tuat.ac.jp (Corresponding author: W. K.)

B. Yu-Jung Lu
Research Center for Applied Sciences, Academia Sinica, Taipei 11529, Taiwan.

C. Satoshi Iwamoto
Research Center for Advanced Science and Technology, The University of Tokyo, 4-6-1 Komaba, Meguro-ku, Tokyo 153-8904, Japan

Funding: This work was supported by JSPS KAKENHI Grant Number JP24H02232.

Keywords: Non-Hermitian physics, Unidirectionality, Infrared region, Multilayer,

Unidirectional electromagnetic absorption provides a powerful approach for controlling light and heat, yet broadband realization in the infrared spectral region remains experimentally unexplored. Here, we report a non-Hermitian multilayer structure that enables robust broadband infrared unidirectional absorption. By combining low- and high-loss materials and engineering their thicknesses using the transfer-matrix formulation, the structure exhibits nearly perfect absorption spectrally matched to the blackbody radiation at 373 K under forward illumination, while suppressing backward absorption below 30%. Spectral analysis indicates that the observed unidirectionality originates from non-Hermitian physics near an exceptional point. Notably, broadband unidirectional absorption is achieved even without strict exceptional-point condition. This indicates that the observed unidirectionality is governed by the combination effect of loss distribution and optical interference, rather than a singular condition, ensuring robustness against film thickness variations. Furthermore, thermal shielding experiments demonstrate that the structure enables unidirectional control of thermal radiation, resulting in a temperature difference of up to 21 ℃ between forward and backward configurations. These results establish a robust strategy for broadband directional control of infrared radiation, with potential applications in passive thermal management, including thermal smart windows and infrared heat-shielding devices.

## 1. Introduction

Non-Hermitian physics has emerged as a powerful platform for realizing light–matter interactions, enabling phenomena such as optical unidirectionality [1-13] and the non-Hermitian skin effect. These distinctive optical phenomena open promising opportunities for applications in the visible and infrared (IR) spectral regions, including optical isolators, and thermal radiation control.[4, 14-16] Beyond photonic platforms, non-Hermitian concepts have recently been extended to thermal [17-20], acoustic [21, 22], and electrical systems [23], where they establish a unified framework for controlling energy, heat, and charge transport under nonequilibrium conditions.

Representative demonstrations of non-Hermitian unidirectionality have been reported in the visible region. For example, Feng et al. demonstrated a multilayer structure exhibiting optical unidirectionality at 532 nm, arising from an abrupt phase transition at an exceptional point.[2] In this context, unidirectionality refers to the directional dependence of the optical responses, where the multilayer exhibits different reflection and absorption characteristics under forward and backward illumination. This work also provided direct visualization through reflection imaging under forward and backward illumination.

Non-Hermitian unidirectionality has also been explored beyond the visible regime. Experimental demonstrations include unidirectional reflectionless propagation in metasurfaces [24] and asymmetric reflection and radiation in photonic systems.[25, 26] These studies indicate that non-Hermitian physics can induce directional optical responses at wavelengths longer than the visible regime, including the infrared. However, these demonstrations are typically limited to narrow spectral ranges.[27] Broadband unidirectional absorption in the IR region based on asymmetric loss-induced non-Hermitian multilayer structure, particularly matched to blackbody radiation, remains experimentally unexplored. In addition, practical implementations require robustness against geometric deviations and fabrication imperfections, which has not been sufficiently addressed in previous studies.[28] To address these issues, we propose a design strategy for achieving for broadband IR unidirectional absorption based on two principles, (i) the selection of high-loss materials exhibiting a large imaginary part of the permittivity in the IR region, and (ii) the engineering of asymmetric loss distribution across a multilayer structure to achieve unidirectional absorption through thickness optimization using the transfer matrix method.

In this study, we design and demonstrate a non-Hermitian multilayer structure that enables broadband infrared unidirectional absorption and is spectrally matched to blackbody radiation using the transfer-matrix formulation. Interestingly, the designed multilayer exhibits significant unidirectional absorption even without satisfying the exact exceptional-point condition, indicating that broadband directional absorption can be robustly achieved in the proposed multilayer structure. We also demonstrate its thermal radiation shielding functionality, highlighting its potential for practical thermal management applications such as thermal smart materials.

## 2. Theory

We considered a multilayer structure composed of alternating low-loss and high-loss materials to realize unidirectional absorption in the IR region. The alternating arrangement of high-loss and low-loss layers enables spatial control of the loss distribution across the multilayer, which is essential for achieving directional energy dissipation. In this configuration, the low-loss layers function as dielectric spacers that mediate optical interference between the high-loss layers, while the high-loss layers provide strong absorption. This interplay between interference and asymmetric loss distribution leads to unidirectional absorption in the multilayer structure. This mechanism highlights that unidirectional absorption arises not from geometrical asymmetry alone but from the engineered interplay between loss distribution and interference.

In this study, calcium fluoride ($CaF_2$) and bismuth (Bi) were selected as representative low-loss and high-loss materials, respectively. Supporting Information **Figure S1(a, b)** shows the refractive index *n* and extinction coefficient *k* of $CaF_2$ and Bi in the IR region. In this study, we designed a non-Hermitian structure consisting of four layers.[29, 30]

To achieve unidirectionality of the multilayer structure, the thicknesses of the four layers were designed using the electromagnetic transfer-matrix formulation.[31] The relationship between the electromagnetic fields at the incident side and the substrate side is expressed as

$$\begin{pmatrix} \boldsymbol{E_0} \\ \boldsymbol{H_0} \end{pmatrix} = \mathbf{M} \begin{pmatrix} \boldsymbol{E_{N+1}} \\ \boldsymbol{H_{N+1}} \end{pmatrix}, (1),$$

where $\mathbf{M}$ is the total transfer matrix of the multilayer structure, given by

$$\mathbf{M} = \prod_{j=1}^{N} \begin{pmatrix} \boldsymbol{cos\beta_j} & \boldsymbol{isin\frac{\beta_j}{q_j}} \\ \boldsymbol{iq_j sin\beta_j} & \boldsymbol{cos\beta_j} \end{pmatrix},\ \boldsymbol{j = 1, 2, 3, \cdots N}, (2)$$

This formulation provides a compact description of multiple reflections, interference, and absorption within the multilayer, enabling efficient analysis of direction-dependent optical responses.

Here, $\boldsymbol{\beta_j}$ is the phase thickness of the *j*-th layer defined as

$$\boldsymbol{\beta_j = k_0 n_j d_j cos\theta_j}, (3)$$

The optical admittance $\boldsymbol{q_j}$ depends on the polarization of the incident light.

$$\boldsymbol{q_j}(\mathbf{TE}) = \boldsymbol{n_j cos\theta_j},\ \boldsymbol{q_j}(\mathbf{TM}) = \frac{\boldsymbol{n_j}}{\boldsymbol{cos\theta_j}}, \quad (4)$$

where $\boldsymbol{n_j}$, $\boldsymbol{d_j}$, and $\boldsymbol{k_0}$ are the refractive index, the thickness of the *j*-th layer, and the wavenumber in vacuum.

The forward and backward reflection and transmission coefficients can be calculated using Equations (5-6),

$$\boldsymbol{r_f} = \frac{(\boldsymbol{M^f_{11}} + \boldsymbol{M^f_{12}}\boldsymbol{q_{N+1}})\boldsymbol{q_0} - (\boldsymbol{M^f_{21}} + \boldsymbol{M^f_{22}}\boldsymbol{q_{N+1}})}{(\boldsymbol{M^f_{11}} + \boldsymbol{M^f_{12}}\boldsymbol{q_{N+1}})\boldsymbol{q_0} + (\boldsymbol{M^f_{21}} + \boldsymbol{M^f_{22}}\boldsymbol{q_{N+1}})}, \quad (5)$$

$$\boldsymbol{r_b} = \frac{(\boldsymbol{M^b_{11}} + \boldsymbol{M^b_{12}}\boldsymbol{q_{N+1}})\boldsymbol{q_0} - (\boldsymbol{M^b_{21}} + \boldsymbol{M^b_{22}}\boldsymbol{q_{N+1}})}{(\boldsymbol{M^b_{11}} + \boldsymbol{M^b_{12}}\boldsymbol{q_{N+1}})\boldsymbol{q_0} + (\boldsymbol{M^b_{21}} + \boldsymbol{M^b_{22}}\boldsymbol{q_{N+1}})}, \quad (6)$$

The total transfer matrices are constructed by Equations (7, 8).

$$\boldsymbol{M^f} = \boldsymbol{M_N} \cdot \cdots \cdot \boldsymbol{M_2 M_1}. \quad (7)$$

$$\boldsymbol{M^b} = \boldsymbol{M_1 M_2} \cdot \cdots \cdot \boldsymbol{M_N}. \quad (8)$$

As the non-Hermitian structure contains high-loss materials, the refractive index becomes complex ($n + \mathrm{i}\boldsymbol{\kappa}$) in the lossy layers, leading to energy dissipation during wave propagation and breaking Hermiticity. In Hermitian systems, energy is conserved during wave propagation, whereas the introduction of loss results in non-conservative optical behavior. As a result, the transfer matrices become non-Hermitian, causing different propagation behavior for forward and backward directions. This leads to direction-dependent field penetration and energy dissipation, governed by the engineered loss distribution and optical interference, resulting in asymmetric reflection and absorption. Using this framework, the multilayer thicknesses were optimized to maximize directional asymmetry through loss coupling and interference engineering.

**3. Results and Discussions**

**Figure 1**(a) presents a schematic of the four-layer structure. The total thickness of the four-layer structure, as well as the thickness of each individual layer, was designed to match the blackbody radiation spectra at different temperatures. **Figures 1**(b-f) show the calculated absorption spectra of the four-layer structure under forward (red line) and backward (black line) illumination. The absorption spectra were calculated using $\boldsymbol{A} = \mathbf{1} - \boldsymbol{T} - \boldsymbol{R}$, where $\boldsymbol{T}$ and $\boldsymbol{R}$ denote the transmission and reflection, respectively (**Figure S2**). The structures were designed to match the blackbody radiation spectra at (b) 1000 K, (c) 500 K, (d) 373 K, (e) 300 K, and (f) 273 K. The corresponding designed total film thicknesses are (b) 337 nm, (c) 682 nm, (d) 978 nm, (e) 1276 nm, and (f) 1570 nm.

All samples exhibited unidirectional absorption at distinct wavelengths. The absorption peak systematically shifted with increasing total film thickness, arising from thickness-dependent constructive and destructive interference in the multilayer. As a result, the peak position can be tuned to match the blackbody radiation spectrum at different temperatures, indicating that the four-layer structure can function as a thermal radiation shielding structure over a wide temperature range.

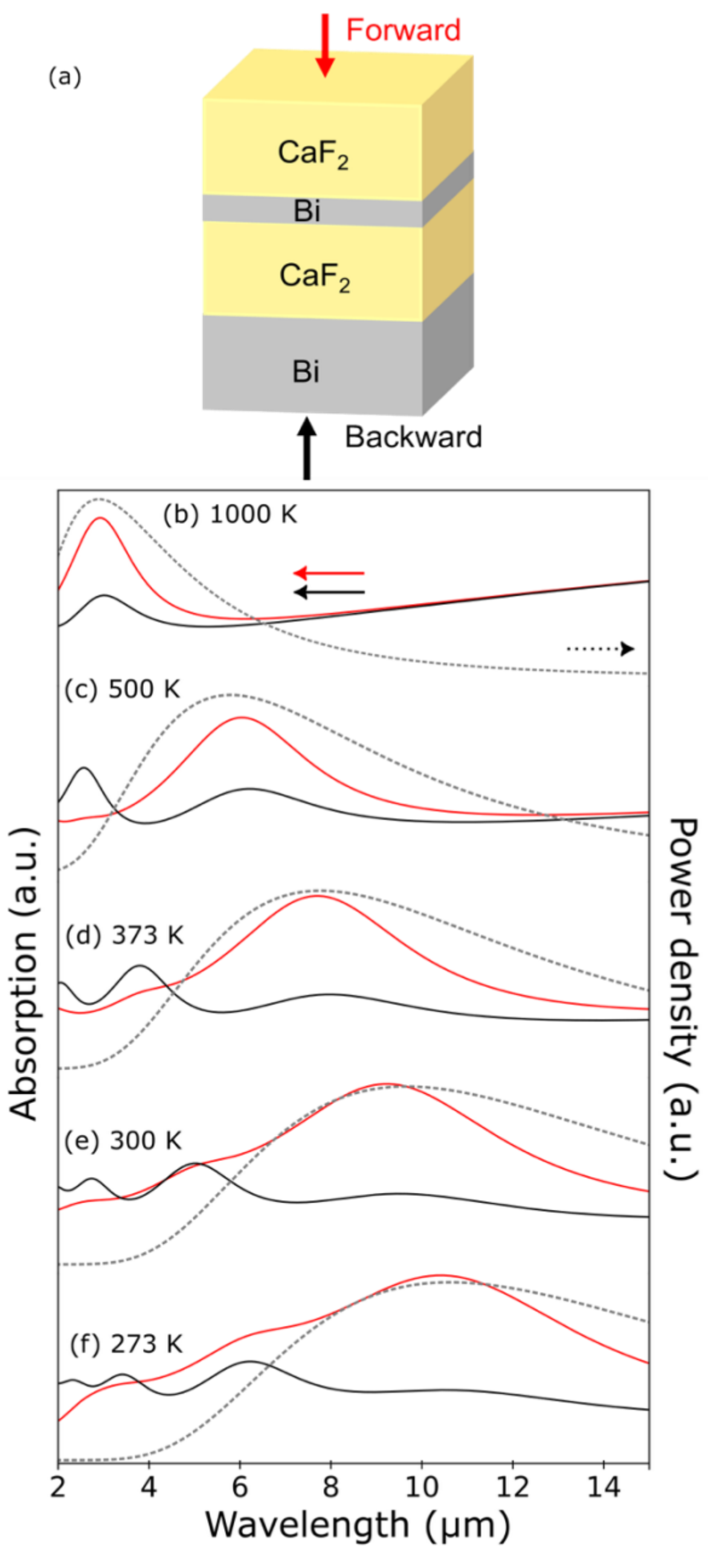

**Figure 1.** (a) Schematic of the four-layer structure. (b-f) Calculated absorption spectra of the four-layer structure designed to match the blackbody radiation spectra at (b) 1000 K, (c) 500 K, (d) 373 K, (e) 300 K, and (f) 273 K, under forward (red line) and backward (black line)

illumination. The designed total film thicknesses were (b) 337 nm, (c) 682 nm, (d) 978 nm, (e) 1276 nm, and (f) 1570 nm. The dashed lines represent blackbody radiation spectra.

**Figure 2**(a) shows the calculated absorption spectra of the four-layer structure with a total thickness of 978 nm under forward (red line) and backward (black line) illumination. The structure consists of $CaF_2$ (top layer)/Bi/$CaF_2$/Bi (bottom layer) layers with thicknesses of 300 nm, 78 nm, 300 nm, and 300 nm, respectively. The structure exhibits the absorption values of 92% and 40% under forward and backward illumination, respectively, at a wavelength of 7.57 μm, corresponding to a modulation depth (MD) of 52.0%. Here, the modulation depth is defined as the difference between forward and backward absorption values.

$\boldsymbol{Modulation\ depth}\ (\mathbf{MD}) = \boldsymbol{A_f - A_b}$, Eq. (9).

The absorption spectrum showed good agreement with the blackbody radiation spectrum at 373 K (gray dashed line). The calculated absorbed power density is $4.9 \times \mathbf{10^8}$ W $m^{-3}$. This value represents the amount of thermal radiation absorbed by the multilayer structure per unit volume. It was calculated by integrating the spectral overlap between the absorption spectrum and the blackbody radiation spectrum at 373 K over the wavelength range from 3 to 13 μm.

Although the multilayer structure was not explicitly designed to satisfy an exceptional-point condition, the designed structure exhibited strong directional absorption arising from asymmetric loss coupling and interference engineering. To verify that the observed unidirectional absorption originates from non-Hermitian physics, we calculated the absorption spectrum of a symmetric four-layer structure composed of Bi layers with identical thickness, as shown in **Figure S3(a)**. The symmetric structure does not exhibit significant unidirectionality (**Figure S3(b)**). These results suggest that the asymmetric loss distribution is essential for realizing unidirectional absorption. Since the directional absorption appears only in the presence of asymmetric loss, the observed unidirectional behavior is consistent with a non-Hermitian optical response of the multilayer structure.

To further discuss the non-Hermitian characteristics of the optimized multilayer structure and examine whether the designed structure operates in the vicinity of an exceptional point, we calculated the generalized total power and its wavelength derivative as functions of wavelength (**Figure 2**(b)). The generalized total power $\boldsymbol{S_n}$ can be calculated using Equation (10),

$\boldsymbol{S_n = T \pm \sqrt{R_f R_b}}$, (10)

where $\boldsymbol{T} = |\boldsymbol{t}|^2$, and $\boldsymbol{R}_\mathrm{f} = |\boldsymbol{r}_\mathrm{f}|^2$ and $\boldsymbol{R}_\mathrm{b} = |\boldsymbol{r}_\mathrm{b}|^2$ are the reflectance values in the forward and backward directions.[2] The generalized total power provides a measure of energy dissipation in the multilayer structure.[2]

Near a non-Hermitian degeneracy, two eigenmodes become strongly coupled, causing abrupt phase variation in the optical response. Such rapid phase change indicates dramatic changes in the interference conditions and energy dissipation within the multilayer structure.[32, 3334] s. Therefore, the combined analysis of the generalized total power and its wavelength derivative is useful for identifying spectral regions where the system operates near an exceptional point.

The generalized total power spectrum shows a minimum value at 7.6 μm, corresponding to the peak wavelength of the absorption under forward illumination. Simultaneously, the derivative exhibits an abrupt phase change at this wavelength, suggesting that the optical interference significantly changes around the peak wavelength. This behavior is consistent with strong mode coupling near an exceptional point, although the system does not exactly satisfy the exceptional-point condition.

**Figure 2**(c) presents the magnetic field distributions of the four-layer structure under forward and backward illumination at 7.5 μm. Under forward illumination, the incident magnetic field penetrates into the bottom layer, leading to nearly perfect absorption. In contrast, under backward illumination, the incident magnetic field penetrates only into the first three layers, where the incident light is predominantly reflected and escapes back into free space. These results indicate that the asymmetric magnetic field penetration is consistent with the observed unidirectional absorption of the four-layer structure.

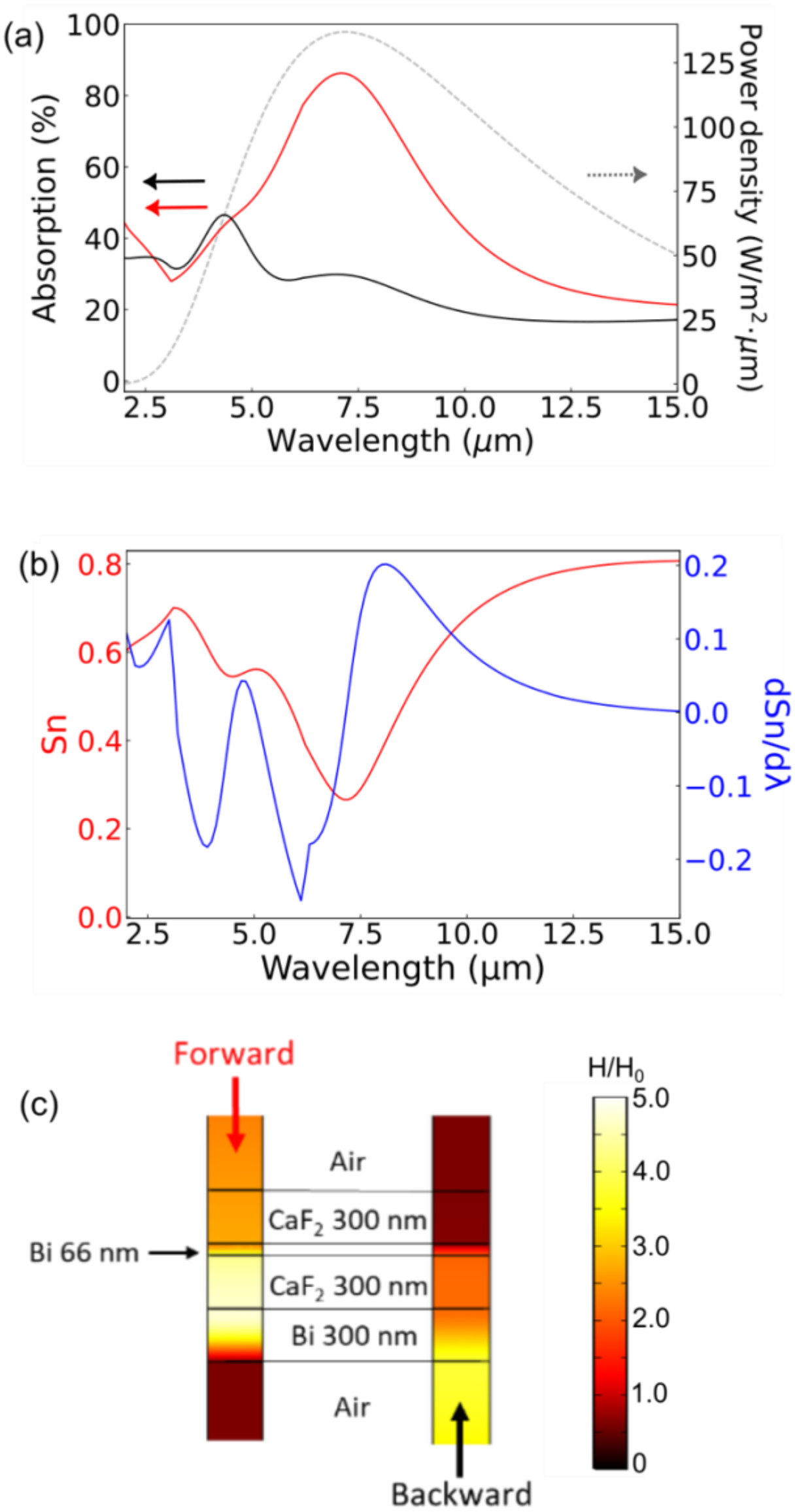

**Figure 2.** (a) Calculated absorption spectra of the four-layer structure with a total thickness of 978 nm under forward (red line) and backward (black line) illumination. The gray dashed line presents the blackbody radiation spectrum at 373 K. (b) Calculated spectra of the generalized total power and its wavelength derivative as functions of wavelength for the four-layer structure. (c) Magnetic field distributions of the four-layer structure under forward and backward illumination at 7.5 μm.

To realize broadband infrared unidirectional absorption, we fabricated the four-layer structure on a Cu substrate using thermal evaporation. As this work represents the first experimental demonstration of the proposed non-Hermitian multilayer structure, the forward and backward structures were fabricated separately on two Cu substrates to clearly evaluate their respective characteristics.

**Figure 3**(a) shows a photograph of the fabricated four-layer structure with a total thickness of 978 nm. The left sample corresponds to the forward configuration ($CaF_2$ (top layer)/Bi/$CaF_2$/Bi (bottom layer)), while the right sample corresponds to the backward configuration (Bi (top layer)/$CaF_2$/Bi/$CaF_2$ (bottom layer)). The two samples show different colors, indicating distinct reflection characteristics even in the visible region.

**Figure 3**(b) presents a cross-sectional SEM image of the forward structure. **Table 1** summarizes the designed and measured film thicknesses, which were evaluated from the SEM images. The measured film thicknesses of the forward structure are $302 \pm 7.1$ nm (top $CaF_2$ layer), $72.3 \pm 9.9$ nm (Bi layer), $309 \pm 8.5$ nm ($CaF_2$ layer), and $168 \pm 4.3$ nm (bottom Bi layer), respectively. For the backward structure, the corresponding thicknesses are $238 \pm 2.8$ nm (top Bi layer), $261 \pm 2.9$ nm ($CaF_2$ layer), $78.0 \pm 1.5$ nm (Bi layer), and $265 \pm 4.3$ nm (bottom $CaF_2$ layer). The difference in thickness between the bottom Bi layer of the forward structure and the top Bi layer of the backward structure arises from the fabrication process. The third ($CaF_2$), second (Bi), and first (top $CaF_2$) layers of the forward structure were deposited simultaneously with the fourth ($CaF_2$), third (Bi), and second ($CaF_2$) layers of the backward structure. Therefore, these layers have almost identical thicknesses. In contrast, the fourth (bottom) layer (Bi) of the forward structure and the top layer (Bi) of the backward structure were deposited separately. As a result, a slight difference in thickness was observed due to the fabrication process.

**Table 1.** Comparison of the designed and measured film thicknesses of the forward and backward structures. (Measured number $n = 5$)

| Designed film thicknesses of the forward structure (nm) | Measured film thicknesses of forward structure (nm) | Designed film thicknesses of the backward structure (nm) | Measured film thicknesses of the backward structure (nm) |
|---|---|---|---|
| $CaF_2$ 300 nm | $302 \pm 7.1$ | Bi 300 nm | $238 \pm 2.8$ |
| Bi 78 nm | $72.3 \pm 9.9$ | $CaF_2$ 300 nm | $261 \pm 2.9$ |
| $CaF_2$ 300 nm | $309 \pm 8.5$ | Bi 78 nm | $78.0 \pm 1.5$ |
| Bi 300 nm | $168 \pm 4.3$ | $CaF_2$ 300 nm | $265 \pm 4.3$ |

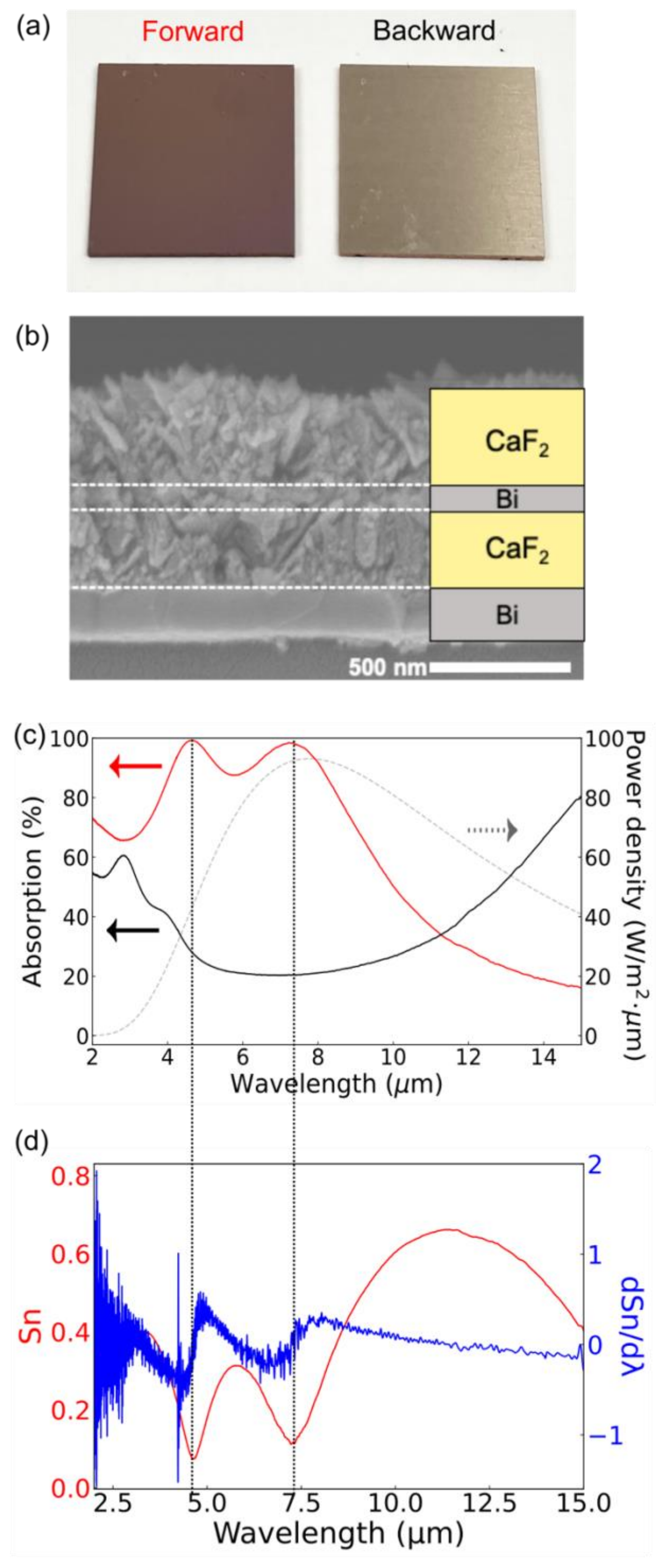

**Figure 3.** (a) Photograph of the fabricated four-layer structure with a designed total thickness of 978 nm. The left and right regions correspond to film stacks of $CaF_2$ (top layer)/Bi/$CaF_2$/Bi (bottom layer)/Cu (substrate) and Bi (top layer)/$CaF_2$/Bi/$CaF_2$ (bottom layer)/Cu (substrate), respectively. (b) Cross-sectional SEM image of the forward structure with the designed total thickness of 978 nm. (c) Measured absorption spectra of the sample under forward (red line)

and backward (black line) illumination. The dashed line presents the blackbody radiation spectrum at 373 K. (d) Comparison of the generalized total power spectrum and its wavelength derivative calculated from the measured absorption spectra of the four-layer structure.

**Figure 3**(c) presents the measured absorption spectra of the forward (red line) and backward (black line) structures. A reflection measurement unit with an oblique incidence angle of 5° was used for the reflection measurements, and the incident light was unpolarized in all measurements. A bare Cu substrate was used as a reference sample.

Nearly perfect absorption values of 99.1% and 98.4% were observed at wavelengths of 4.7 and 7.2 μm, respectively. In contrast, the absorption values under backward illumination at these wavelengths were 27.1% and 20.4%, indicating a modulation depth of 52% at 4.7 μm between forward and backward illumination. **Figure 3**(c) exhibits two peaks at 4.7 and 7.2 μm, whereas the calculated absorption spectrum shows a single peak at 7.6 μm. This discrepancy is attributed to variations in film thickness within the multilayer structures. To verify this, we calculated absorption spectra for thinner (Configuration 1) and thicker (Configuration 2) structures. The corresponding film thicknesses are 300 nm (top $CaF_2$ layer)/26 nm (Bi)/300 nm ($CaF_2$)/163 nm (bottom Bi layer) and 300 nm (top $CaF_2$ layer)/66 nm (Bi)/300 nm ($CaF_2$)/300 nm (bottom Bi layer), respectively. These configurations were defined based on the standard deviations of the film thicknesses estimated from the SEM measurements. The calculated peaks of these configurations reproduce the two peaks observed in the measured spectrum, confirming their origin (**Figure S4**). The measured absorption spectrum also agrees well with the blackbody radiation spectrum at 373 K (gray dashed line), indicating that the structure exhibits unidirectional absorption for thermal radiation.

Although the polarization of the incident light changes the optical admittance $\boldsymbol{q_j}$, (Equation (4)), numerical calculations reveal that the absorption characteristics are insensitive to both polarization and angle of incidence (**Figure S5**). These results indicate that the 5° oblique incidence used in the measurements does not significantly influence the observed broadband infrared unidirectionality.

To further confirm that this unidirectionality originates from non-Hermitian physics, we calculated the spectra of the generalized total power and its wavelength derivative (**Figure 3**(d)). The minimum $\boldsymbol{S_n}$ value and an abrupt phase change variation in the derivative spectrum are observed at 4.7 μm, indicating that the system operates near, but not exactly at, an exceptional point. Even under this condition, pronounced unidirectionality is observed, suggesting that the directional response is robust against slight deviations from the ideal non-Hermitian design.

This result represents the first experimental demonstration of broadband infrared unidirectionality enabled by non-Hermitian physics.

We evaluated the thermal shielding performance of the four-layer structure. **Figure 4**(a) presents the experimental setup, where the forward and backward structures were placed facing a hot plate set at 500 °C with a gap of 10 cm, and the rear side temperatures were monitored using an IR camera. The forward structure exhibits a rear-side temperature of 133 °C, whereas the backward structure shows a lower temperature of 112 °C due to partial reflection of incident thermal radiation, resulting in a temperature difference of 21 °C. These results suggest that the structure enables unidirectional control of thermal radiation, highlighting the potential of the structure as an effective thermal shielding material.

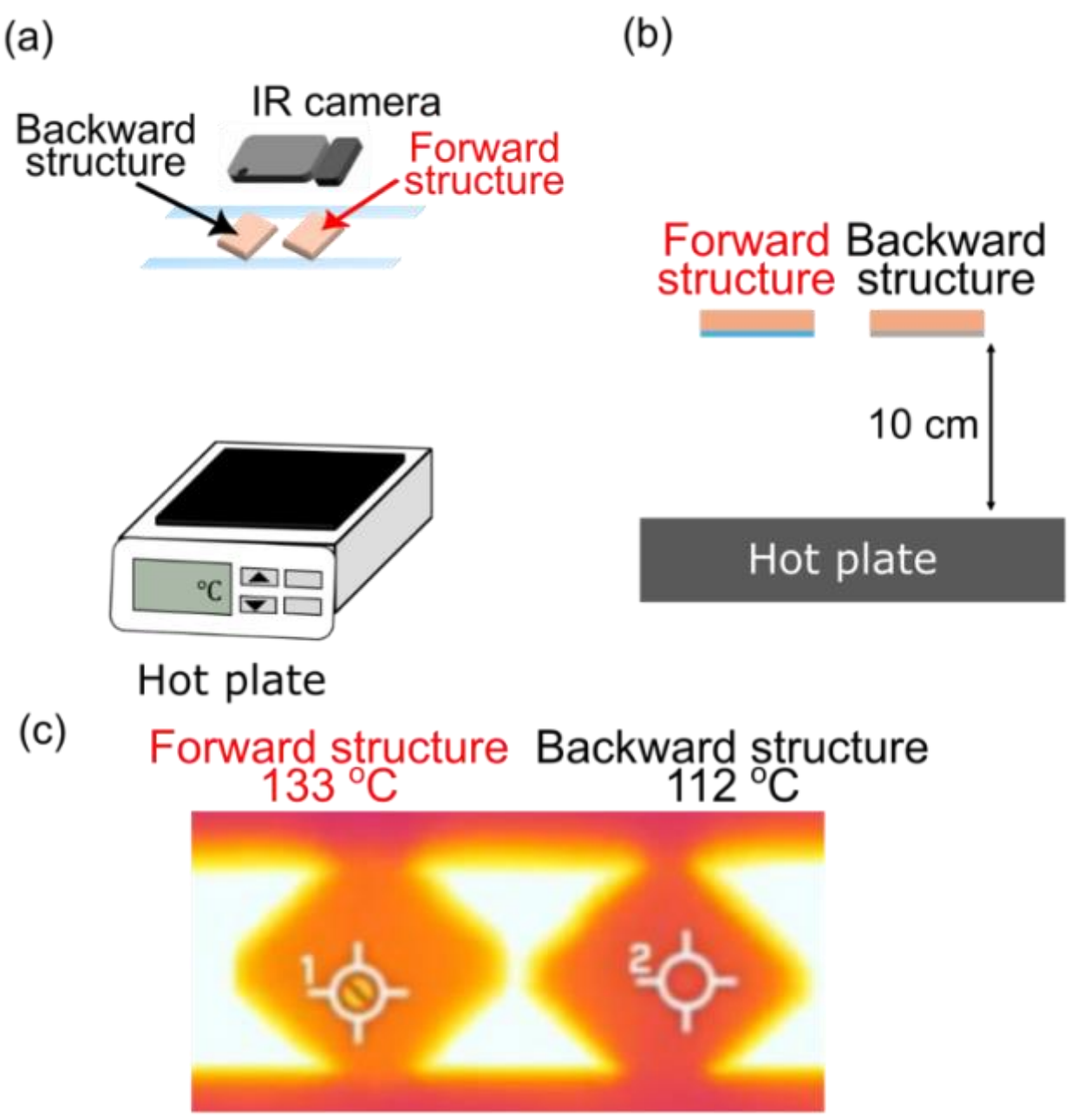

**Figure 4.** (a) Experimental setup for observing the thermal shielding effect of the four-layer structure. (b) Cross-sectional schematic of the experimental setup. (c) Infrared (IR) camera images of the non-Hermitian structures with forward and backward film-stack configurations.

## 4. Conclusion

In summary, we demonstrated a non-Hermitian four-layer structure exhibiting broadband infrared unidirectional absorption. By engineering the layer thicknesses using the transfer-matrix formulation, the forward absorption is spectrally matched to blackbody radiation at different temperatures, while the backward absorption is strongly suppressed. The fabricated structure achieves a modulation depth of 53%.

Analysis of the generalized total power spectrum and its wavelength derivative indicates that the system operates near, but not exactly at, an exceptional point. Despite this, pronounced unidirectionality is maintained, demonstrating robustness against structural deviations.

This work establishes a robust strategy for broadband unidirectional control of infrared radiation using non-Hermitian multilayer structures, with potential applications in passive thermal management, including smart windows and infrared heat-shielding devices.

**5. Experimental Section**

A cleaned Cu substrate with a thickness of 0.3 mm was used as the base for sample fabrication after standard cleaning procedures. The non-Hermitian four-layer structure, consisting of Bi and $CaF_2$ layers, was deposited by thermal evaporation (SVC-7TS 80, Sanyu Electron Corporation). The layer thicknesses were designed using the transfer-matrix formulation to achieve near-perfect absorption under forward illumination, while matching the spectral profile to the blackbody radiation at the target temperature. The thickness of each deposited layer was determined from cross-sectional observations using a scanning electron microscope (Hitachi High-Technologies Corporation, SU8010). Infrared reflection spectra were measured using Fourier transform infrared (FTIR) spectroscopy (JASCO Corporation, FT/IR-4X) at an incident angle of 5°. Thermal images were acquired using an infrared camera (FLIR, FLIR One Pro).

**Data Availability Statement**

The data supporting the findings of this study are available from the corresponding author upon reasonable request. Additional data related to this work are provided in the Supporting Information.

**Supporting Information**

Supporting Information is available from the Wiley Online Library or from the author.

Table of Contents

**Robust Broadband Infrared Unidirectional Absorption Enabled by a Non-Hermitian Multilayer**

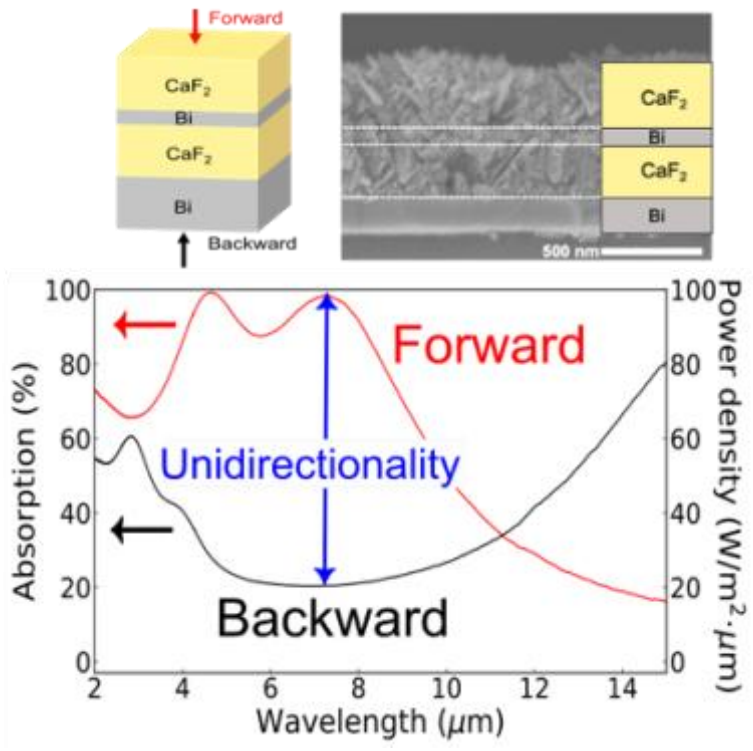

A non-Hermitian multilayer structure enabling broadband infrared unidirectional absorption is demonstrated through the asymmetric distribution of high-loss materials and interference engineering. The designed structure exhibits nearly perfect forward absorption spectrally matched to blackbody radiation, while strongly suppressing backward absorption. This robust approach provides a new strategy for directional thermal radiation control and passive thermal management.

## Supporting Information

**Robust Broadband Infrared Unidirectional Absorption Enabled by a Non-Hermitian Multilayer**

*Ayaha Yamamoto[A], Ganbat Batorgil[A], Yu-Jung Lu[B], Satoshi Iwamoto[C], and Wakana Kubo[A]**

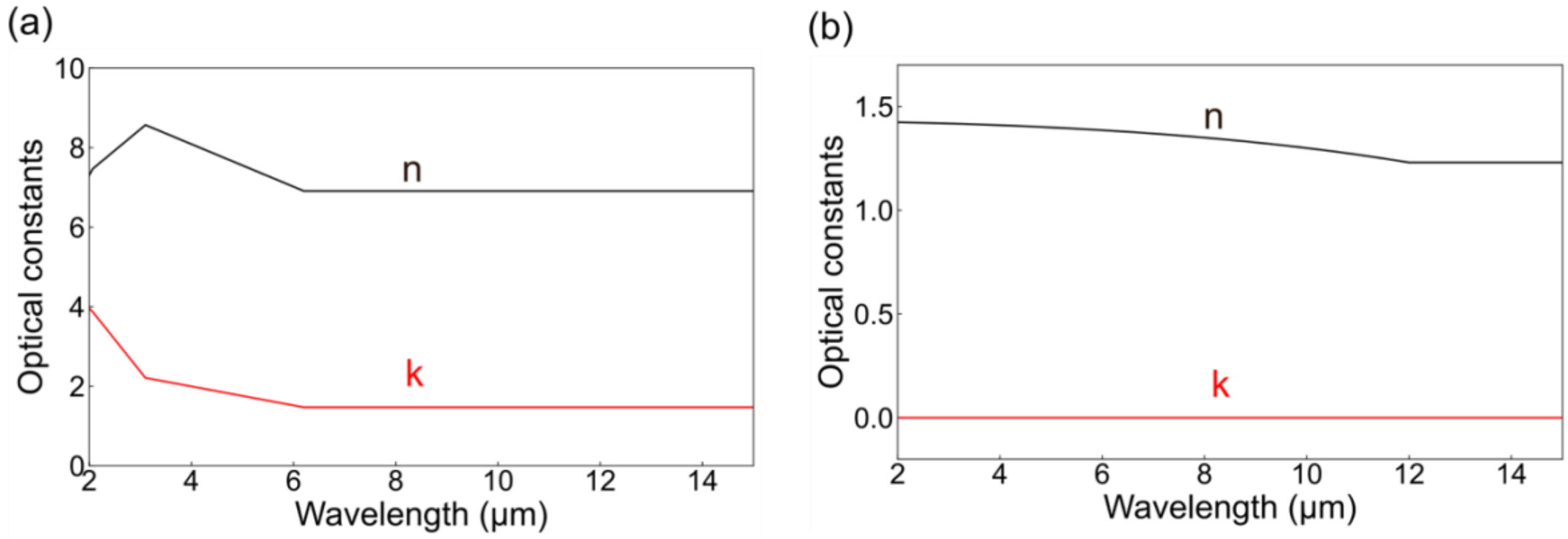

**Figure S1.** Spectral refractive index (n; black line) and extinction coefficient (k; red line) of (a) Bi and (b) $CaF_2$ used in the transfer-matrix calculations.

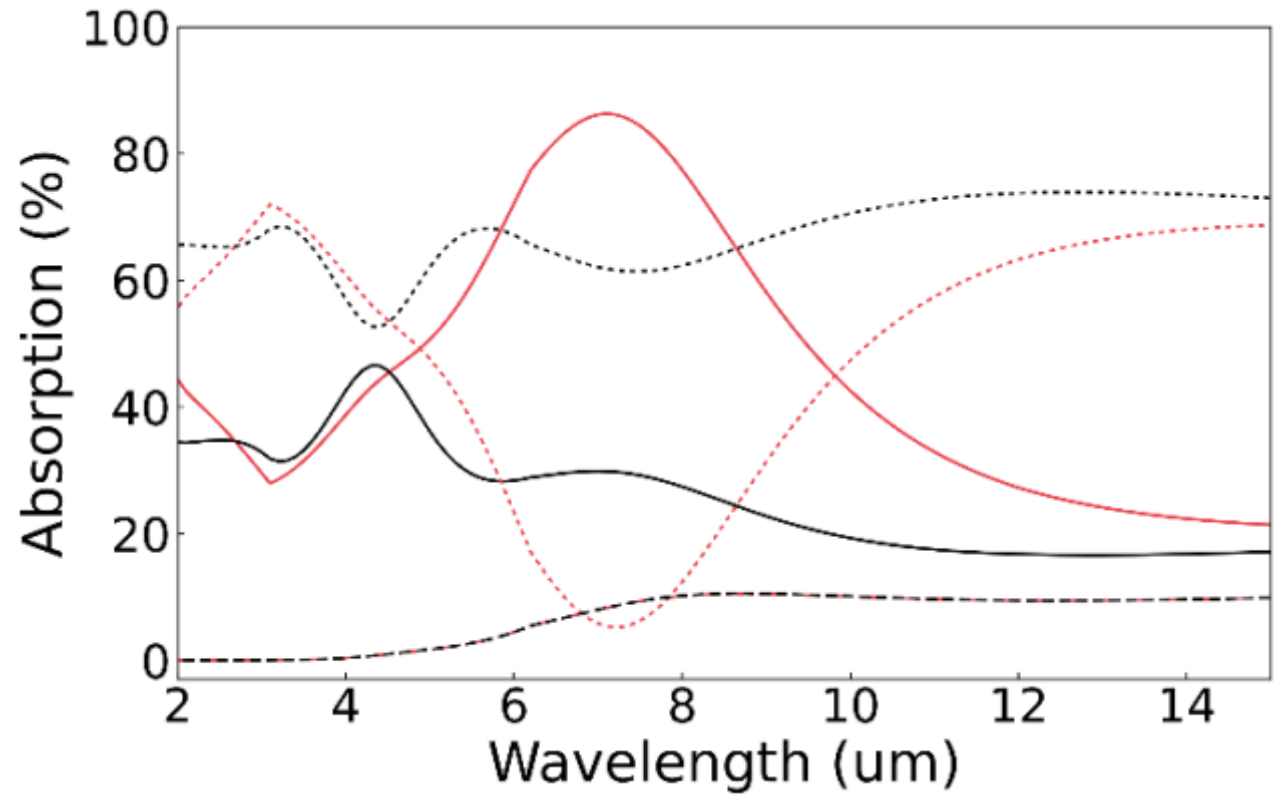

**Figure S2.** Comparison of the calculated absorption (solid lines), reflection (dotted lines), and transmission (dashed lines) spectra of the four-layer structure. The red and black lines present forward and backward illumination, respectively. The transmission spectra (dashed lines) for forward and backward illumination almost coincide and are therefore overlapping.

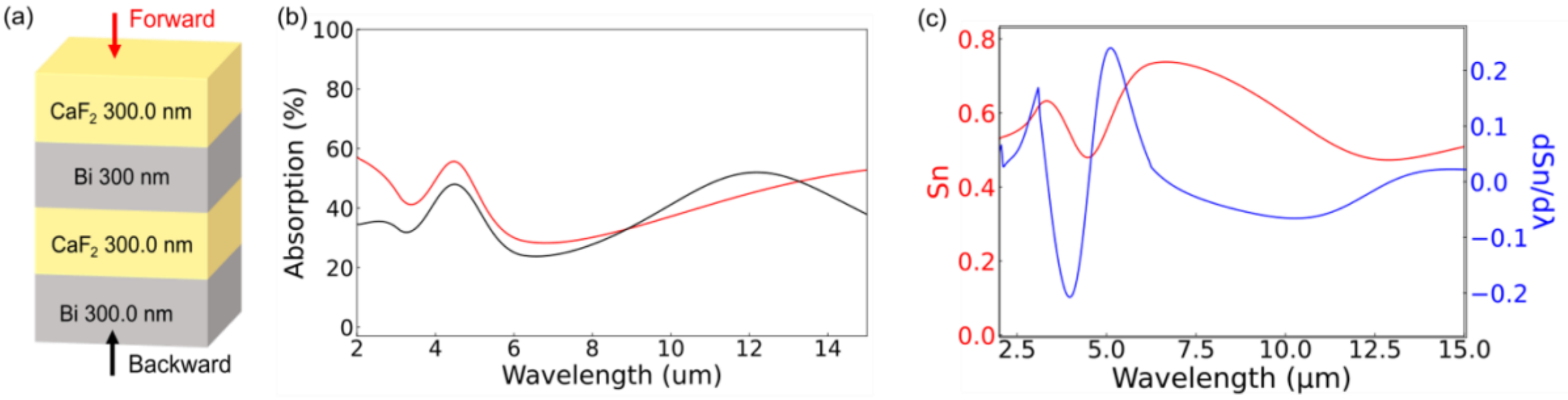

**Figure S3.** (a) Schematic of the symmetric four-layer structure with film stacks of 300 nm (Top $CaF_2$ layer)/300 nm (Bi)/300 nm ($CaF_2$)/300 nm (bottom Bi layer). (b) Calculated absorption spectra under forward (red line) and backward (blue line) illumination. (c) Calculated generalized total power (red line) and its wavelength derivative (blue line) as functions of wavelength for the four-layer structure.

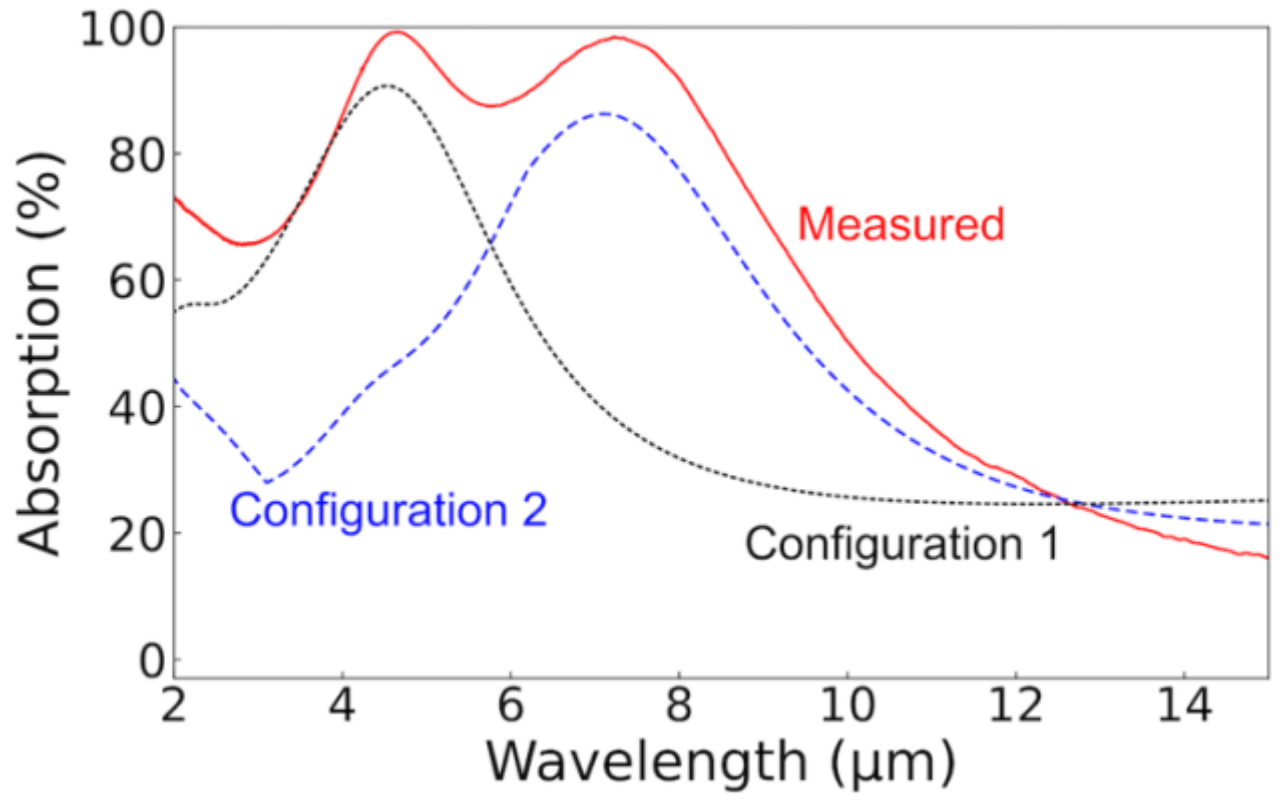

**Figure S4.** Calculated absorption spectra of the four-layer structures with film stacks of 300 nm (Top $CaF_2$ layer)/26 nm (Bi)/300 nm ($CaF_2$)/163 nm (bottom Bi layer) (Configuration 1) and 300 nm (Top $CaF_2$ layer)/66 nm (Bi)/300 nm ($CaF_2$)/300 nm (bottom Bi layer) (Configuration 2), respectively.

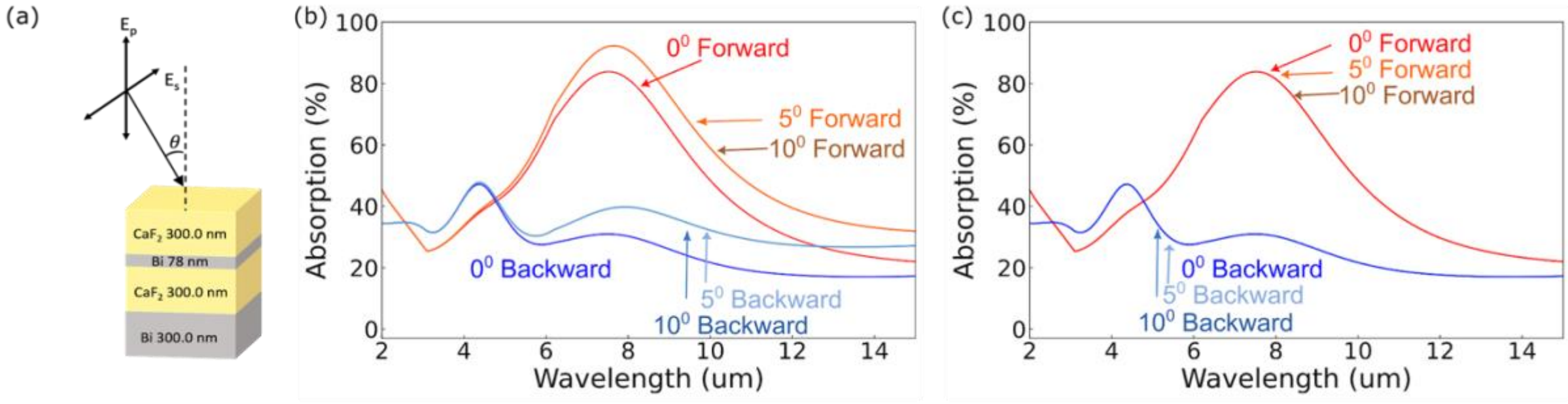

**Figure S5.** (a) Schematic illustration of TE polarization. (b, c) Calculated absorption spectra of the non-Hermitian four-layer structure at different angles of incidence for (b) TE and (c) TM polarizations.